\documentclass[preprints,article,accept,pdftex,moreauthors]{Definitions/mdpi} 

\firstpage{1} 
\makeatletter 
\Title{Hadron Emission and Stopping in Heavy-Ion Collisions: Baryon-Rich Matter to Meson-Dominated Matter}

\TitleCitation{Hadron Emission in Heavy-Ion Collisions}

\usepackage{comment}

\Author{Manuel Lorenz $^{1,2}$\orcidA{}, Christoph Blume $^{2,1,3}$}

\AuthorNames{Manuel Lorenz, Christoph Blume}

\isAPAStyle{%
       \AuthorCitation{Lastname, F., Lastname, F., \& Lastname, F.}
         }{%
        \isChicagoStyle{%
        \AuthorCitation{Lastname, Firstname, Firstname Lastname, and Firstname Lastname.}
        }{
        \AuthorCitation{Lastname, F.; Lastname, F.; Lastname, F.}
        }
}

\address{%
$^{1}$ \quad GSI Helmholtzzentrum für Schwerionenforschung, Darmstadt, Germany; \\
$^{2}$ \quad Institut f\"{u}r Kernphysik, Goethe-Universit\"{a}t, 60438 ~Frankfurt, Germany; \\
$^{3}$ \quad Helmholtz Research Academy Hesse for FAIR (HFHF), GSI Helmholtz Centre for Heavy-Ion Research, Darmstadt, Germany; }

\corres{Correspondence: m.lorenz@gsi.de; Tel.:+49 6159 71 1303}

\abstract{
Today's accelerator facilities used for studies of relativistic heavy-ion collisions cover an energy range spanning over three orders of magnitude, from a few GeV up to a few TeV in center-of-mass energy per nucleon pair ($\sqrt{s_{NN}}$).
We present a systematic overview of hadron emission in heavy-ion collisions across this entire energy range. 
The presented energy excitation functions of the approximated baryon and meson yields at mid-rapidity reflect the interplay between baryon stopping and particle production, both of which evolve continuously with increasing energy. 
At low energies (e.g., SIS18, AGS), strong nuclear stopping leads to high net-baryon densities at mid-rapidity and to the abundant formation of nuclear clusters. 
With increasing $\sqrt{s_{NN}}$, the relative baryon stopping power $\langle \delta y \rangle / y_p$ decreases, and meson production becomes dominant.  
The inelasticity, i.e. the fraction of the initial kinetic energy available converted in inelastic reactions into particle production and dynamics, is found to rise rapidly at low energies and then levels off at values around $0.7 - 0.8$.
While at low energies up to $\sim 10$~GeV this available energy seems to be shared by equal amount between the production of new particles and the dynamics of the system, as well as radiation, the latter part starts to dominates at higher energies.
}


\begin{document}


\section{The Landscape of Relativistic Heavy-Ion-Collisions}\label{Intro} 

Systematic studies of relativistic heavy-ion collisions in the laboratory started almost 50 years ago, in the early 1970s at the Lawrence Berkeley National Laboratory (USA) and at the Joint Institute of Nuclear Research (Dubna, Russia). 
The pioneering experiments in Berkeley and Dubna were followed by a program at Berkeley's Bevatron accelerator started in 1984. 
Soon, dedicated programs at higher energies followed at the AGS (Brookhaven) and SPS (CERN). 
From the early 1990s on, these efforts were supplemented by lower-energy beams at SIS18 (GSI, Darmstadt). 

The era of heavy-ion colliders started in 2000 with the commissioning of RHIC at BNL, pushing $\sqrt{s_{NN}}$ up to 200~GeV. 
In 2009, with the start of the LHC at CERN, the energy frontier was extended into the TeV regime. 

Besides the energy increase, detector technology advances played a key role. 
Early detectors relied on ionization in gases or liquids, with analog (optical) readouts such as streamer or bubble chambers \cite{Harris:1987md,NA35:1990teq}. 
While offering large acceptance, they suffered from slow readout and visual analysis requirements. 
Digital detectors such as drift and time-projection chambers (e.g. NA49 \cite{NA49:1999myq}) revolutionized event reconstruction. 

Calorimeters enabled neutral-particle detection (e.g. WA98 \cite{WA98:2000vxl}), while Cherenkov counters provided lepton identification (e.g. NA45 \cite{Tserruya:1993sa}). 
Modern large-scale experiments such as ATLAS \cite{ATLAS:2008xda}, CMS \cite{CMS:2008xjf}, ALICE \cite{ALICE:2008ngc}, and CBM/HADES \cite{CBM:2016kpk,HADES:2009aat} feature layered designs combining tracking, PID, calorimetry, and triggering.

An overview of major heavy-ion facilities and experiments is given in Tab.~\ref{tab:fac}.

\begin{table}[H]
\centering
\begin{tabularx}{\textwidth}{CCCCCC}
\toprule
\textbf{Accelerator} & \textbf{Facility} & \textbf{Years} & \textbf{$\sqrt{s_{NN}}$ [GeV]} & \textbf{Projectile} & \textbf{Experiments}  \\
\midrule
Bevalac & LBNL, Berkeley & 1984--1993 & 2.0--2.4 & C, Ca, Ni, Nb, La, Au & PlasticBall~\cite{GSI-LBL:1982phm}, StreamerChamber~\cite{Harris:1987md}, EOS~\cite{EOS:1994sbm}, DLS~\cite{DLS:1997kbk} \\           
AGS & BNL, Brookhaven & 1986--1994 & 2.8--11.7 & Si, Au & E802/859~\cite{E802:1999hit}, E866/917~\cite{E866:1999ktz}, E810/891~\cite{E810:1995iwp}, E814/877~\cite{E877:1994plr}, E864/941~\cite{E864:2002xhb}, E895/910~\cite{E895:1999ldn} \\ 
SPS & CERN, Geneva & 1986-- & 6.4--17.3 & Be, O, S, Ar, In, Xe, Pb & NA34(HELIOS)~\cite{Masera:1995ck}, NA35~\cite{NA35:1990teq}, NA36~\cite{NA36:1992avc}, NA38/50/60~\cite{NA60:2006ymb}, NA44~\cite{NA44:1996xlh}, NA45(CERES)~\cite{Tserruya:1993sa}, NA49/61~\cite{NA61:2014lfx}, NA52~\cite{NA52NEWMASS:1996uce}, NA57~\cite{NA57:1999xrk}, WA80/93~\cite{WA80:1995xza}, WA85~\cite{WA85:1991nsm}, WA94~\cite{WA94:1995szb}, WA97~\cite{WA97:1999uwz}, WA98~\cite{WA98:2000vxl} \\ 
SIS18 & GSI, Darmstadt & 1992-- & 1.9--2.5 & C, Ar, Ni, Nb, Ag, Au & FOPI~\cite{Ritman:1995td}, KaoS~\cite{KaoS:1992bsd}, HADES~\cite{HADES:2009aat} \\ 
RHIC & BNL, Brookhaven & 2000-- & 3--200 & O, Ar, Au, U & STAR~\cite{STAR:2002eio}, PHENIX~\cite{PHENIX:2003nhg}, BRAHMS~\cite{BRAHMS:2003rdb}, PHOBOS~\cite{PHOBOS:2003aez} \\ 
LHC & CERN, Geneva & 2009-- & up to 5500 & O, Ar, Pb & ALICE~\cite{ALICE:2008ngc}, ATLAS~\cite{ATLAS:2008xda}, CMS~\cite{CMS:2008xjf}, LHCb~\cite{LHCb:2008vvz} \\ 
\bottomrule
\end{tabularx}
\caption{Overview of experiments at major heavy-ion facilities, their operation periods, covered energy range, and projectiles. Ongoing experiments are shown in bold in the source text.}
\label{tab:fac}
\end{table}

\section{General Characteristics of HICs} \label{chap_bulk} 
Within the energy range covered by the present accelerator spanning over three orders of magnitude from a few GeV up to 5~TeV in center-of-mass energy of an initial nucleus-nucleus collision $\sqrt{s_{NN}}$, the kinematic gap between projectile and target rapidity widens from less than one unit in rapidity (u.r.) to almost 18~u.r..  At the same time the passing time~\footnote{$t_{p}=(R_{p}+R_{t})/(\gamma\times v_{beam})$, where $R_{p}$ and $R_{t}$ are the radii of projectile and target nuclei, $\gamma$ is the Lorentz-factor and $v_{beam}$ the beam velocity.} of the colliding nuclei reduces drastically from more than 10~fm/$c$ to values far below 1~fm/$c$.\\ 
As will be discussed in the following, also the kinematic distributions, the yields and the chemical composition of the reaction products emitted in the collision change strongly. Yet, these changes occur remarkably smoothly as a function of the center-of-mass energy $\sqrt{s_{NN}}$.\\
The yield of emitted hadrons in the collision zone results from an interplay between the amount of stopping of the initial nucleons (quarks) in the collision zone, which decreases with increasing $\sqrt{s_{NN}}$, and the production of new particles, which increases with increasing $\sqrt{s_{NN}}$. \\ 

\subsection{Baryon Stopping} 
\label{S_stopping}  

The amount of energy-loss the initial nucleons experience in the reaction is referred to as baryon stopping power in literature and often discussed with respect to its sensitivity to the compressibility of nuclear matter \cite{NA49:1998gaz,BRAHMS:2003wwg,Blume:2007kw,FOPI:2004orn}.  The yield of hadrons emitted from the collision zone therefore results as an interplay between the amount of stopping of the initial nucleons (quarks) and the production of new particles, which increases with increasing $\sqrt{s_{NN}}$. 

The stopping power can be quantified by an analysis of the rapidity distributions of net-baryons, i.e. the difference baryons-antibaryons, as these are determined by the amount of deceleration in the center-of-mass system, caused by the conversion of the initial longitudinal kinetic energy of the nucleons via inelastic reactions into particle production and dynamics.  Experimentally, net-baryon distributions are usually constructed from the measured proton and antiproton (at higher energies) rapidity distributions.  As (anti-) neutrons are usually not measured, their distributions are assumed to be of the same shape as the ones for (anti-)protons, while their yield is estimated by multiplying the (anti-)proton yield by the initial neutron-to-proton ratio.  At higher energies (SPS and above) also heavier baryons, such as $\Lambda$, $\bar{\Lambda}$, $\Xi^-$ and $\bar{\Xi}^+$, are taken into account, if measured.  At lower energies (SIS18) a substantial fraction of the baryon number is contained in light nuclei (mainly d, t, $^3$He, $^4$He), which need to be included in the sum of baryons.

\begin{figure}[tb]
  \centering
  \includegraphics[width=0.95\textwidth]{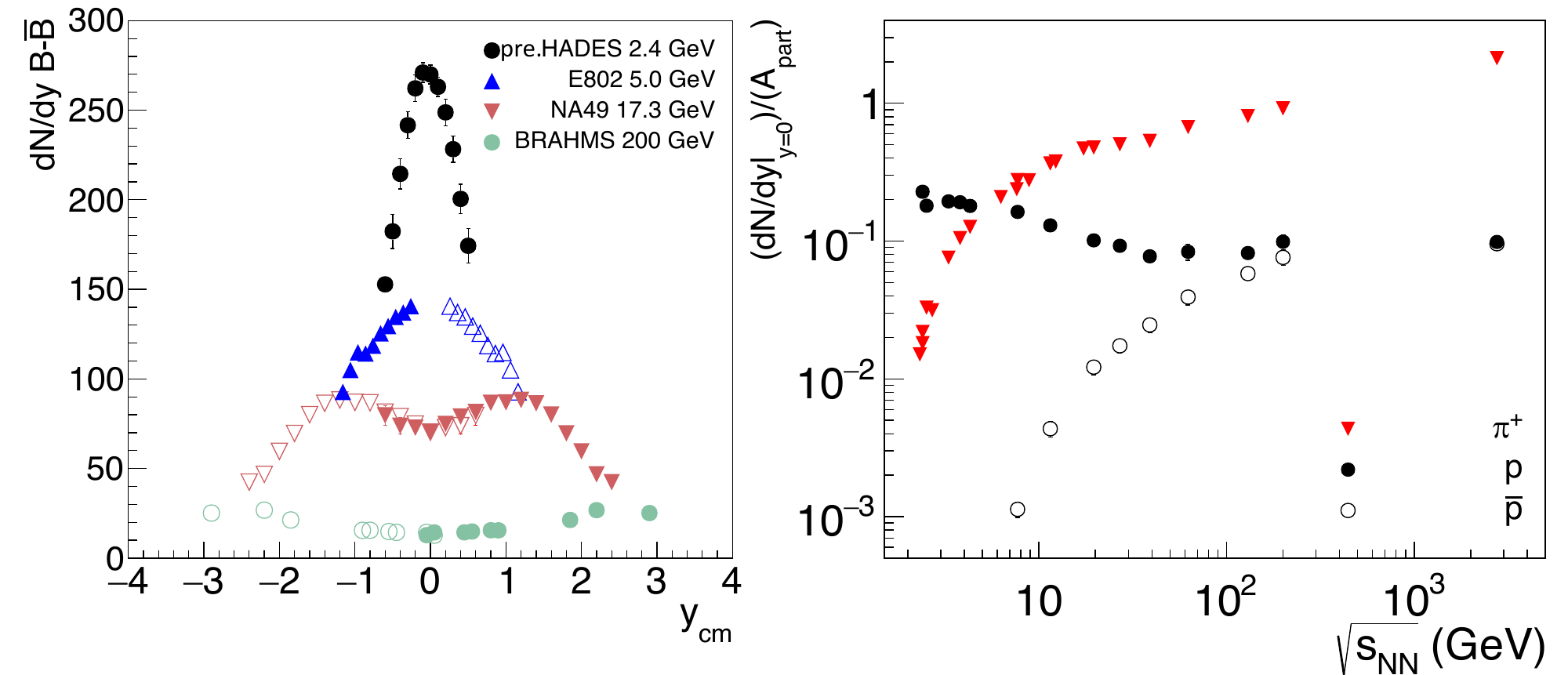}
  \caption[Left: Rapidity distributions of net-baryons at various energies.]{
    Rapidity distributions of net-baryons for central heavy-ion collisions at various collision energies \cite{E-802:1998xum,NA49:1998gaz,BRAHMS:2003wwg} as well as preliminary HADES data. With increasing energy, a valley in the yield distribution of net-baryons around mid-rapidity develops. Right: Energy excitation function of the $\pi^+$, proton and antiproton yields per average participating nucleon $A_{part}$ emitted at mid-rapidity in central Au+Au (Pb+Pb) collisions \cite{FOPI:2006ifg,E-0895:2003oas,NA49:2007stj,NA49:2002pzu,STAR:2008med,STAR:2002hpr,STAR:2017sal,ALICE:2013mez,FOPI:2010xrt,Back:2002ic,E802:1999hit}.}
  \label{fig:net-baryons}
\end{figure}

Figure~\ref{fig:net-baryons} displays exemplary net-baryon rapidity distributions measured in central Au+Au/Pb+Pb collisions at four different center-of-mass energies ($\sqrt{s_{NN}} = 2.4, 5.0, 17.3$ and $200$~GeV) \cite{E-802:1998xum,NA49:1998gaz,BRAHMS:2003wwg}.  A clear evolution of the shape is evident: at low energies (HADES, $\sqrt{s_{NN}} = 2.4$~GeV) a narrow peak around mid-rapidity is observed, which widens towards higher energies (E802, $\sqrt{s_{NN}} = 5.0$~GeV). Around top SPS energies a dip develops around mid-rapidity (NA49, $\sqrt{s_{NN}} = 17.3$~GeV), turning into a shallow valley at RHIC (BRAHMS, $\sqrt{s_{NN}} = 200$~GeV).  As a consequence, the amount of net-baryons present in the collision zone around mid-rapidity at $\sqrt{s_{NN}} = 200$~GeV is more than an order of magnitude smaller than at $\sqrt{s_{NN}} = 2.4$~GeV.  At energies of a few TeV (almost) none of the initial baryons are observed at mid-rapidity  \cite{Andronic:2017pug}.

\subsection{Particle Production} 

Particle production in heavy-ion collisions can be predicted using the principle of maximum entropy \cite{Jaynes:1957zza}. 
The idea of employing statistical methods to describe particle production in nucleon--nucleon or nucleus--nucleus collisions dates back to the late 1940s \cite{Koppe:1949zz,Tawfik:2013tza} and early 1950s \cite{Fermi:1950jd}.
Despite neglecting the details of the collision dynamics, statistical-chemical analyses of hadron yields have become an established tool to characterize particle production with only a few parameters \cite{Andronic:2005yp,Becattini:2005xt,Rafelski:2000by}. 
The multiplicity $M_i$ of hadron species $i$ is proportional to
\[
M_i \;\propto\; \exp\!\left(-\frac{E_i - \sum_j \mu_j Q_{ij}}{T}\right) \, ,
\]

with

\begin{itemize}
  \item $E_i$: energy of the state (for massive particles in the Boltzmann limit approximately $E_i \approx m_i$),
  \item $Q_{ij}$: conserved quantum number $j$ (e.g.\ baryon number, strangeness, electric charge) of particle $i$,
  \item $\mu_j$: corresponding chemical potential, ensuring global conservation of charges,
  \item $T$: Lagrange multiplier from the maximum-entropy principle.
\end{itemize}

Due to the mass term in the exponential factor, the yields of newly produced particles follow a clear mass ordering, with the lightest species, the pions with a mass of about 140~MeV/$c^2$, dominating. 
In most works on statistical-chemical analyses of hadron yields, $T$ is interpreted as the temperature at chemical freeze-out of the system, while baryon stopping is reflected by the increase of the baryo-chemical potential $\mu_B$ toward lower $\sqrt{s_{NN}}$, 
providing a direct mapping in the $T$--$\mu_B$ plane between heavy-ion collisions and the phase diagram of strongly interacting matter.


Experimentally, the yields of pions, protons, and antiprotons have been measured from $\sqrt{s_{NN}}$ of a few GeV up to several TeV. 
For excitation functions presented in this section, we restrict ourselves to mid-rapidity yields, as full $4\pi$ yields are rarely measured at collider energies (for a summary of the available data see the right panel of Fig.~\ref{fig:enetbaryon}). 
The differential yields at mid-rapidity, normalized to the average number of participating nucleons $A_{\text{part}}$, are shown on the right side of Fig.~\ref{fig:net-baryons}. 

The mid-rapidity yield of $\pi^+$ (red triangles) \cite{FOPI:2006ifg,HADES:2020ver,E-0895:2003oas,NA49:2007stj,NA49:2002pzu,STAR:2008med,STAR:2002hpr,STAR:2017sal,ALICE:2013mez}
increases smoothly from below $10^{-2}$ per participant nucleon at a few~GeV to about $2$ at $\sqrt{s_{NN}} = 5$~TeV. 
While the increase is steep at low $\sqrt{s_{NN}}$, the slope flattens in the region between about 5~GeV and 10~GeV.

Due to the higher production threshold and the expected scaling with particle mass, the antiproton yield (open black circles) \cite{STAR:2008med,STAR:2017sal,ALICE:2013mez} starts to rise at larger $\sqrt{s_{NN}}$ compared to that of the pions, and remains significantly lower throughout. 

The proton yield (filled black circles) \cite{FOPI:2010xrt,Back:2002ic,E802:1999hit,STAR:2008med,STAR:2017sal,ALICE:2013mez}, in contrast, reflects two competing effects: baryon stopping into the collision zone and particle production. 
As a consequence, it decreases with increasing $\sqrt{s_{NN}}$ and reaches a minimum around $\sqrt{s_{NN}} \approx 40$~GeV.
At higher energies, the contribution from newly produced protons, predominantly originating from proton--antiproton (quark--antiquark) pair creation, leads to a gentle rise of the proton yield. 
At sufficiently high $\sqrt{s_{NN}}$, the interactions no longer occur between nucleons but rather between partons, mainly gluons and sea quarks, which carry the quantum numbers of the vacuum. 
As a result, the rise of the proton yield becomes increasingly similar to that of the antiprotons. 
At $\sqrt{s_{NN}}=5$~TeV, protons and antiprotons are produced in equal numbers around mid-rapidity, indicating that the transport of initial nucleons and their valence quarks to mid-rapidity is negligible and the number of net-protons in this region approaches zero.

\begin{figure}[tb]
\centering
\includegraphics[width=0.5\textwidth]{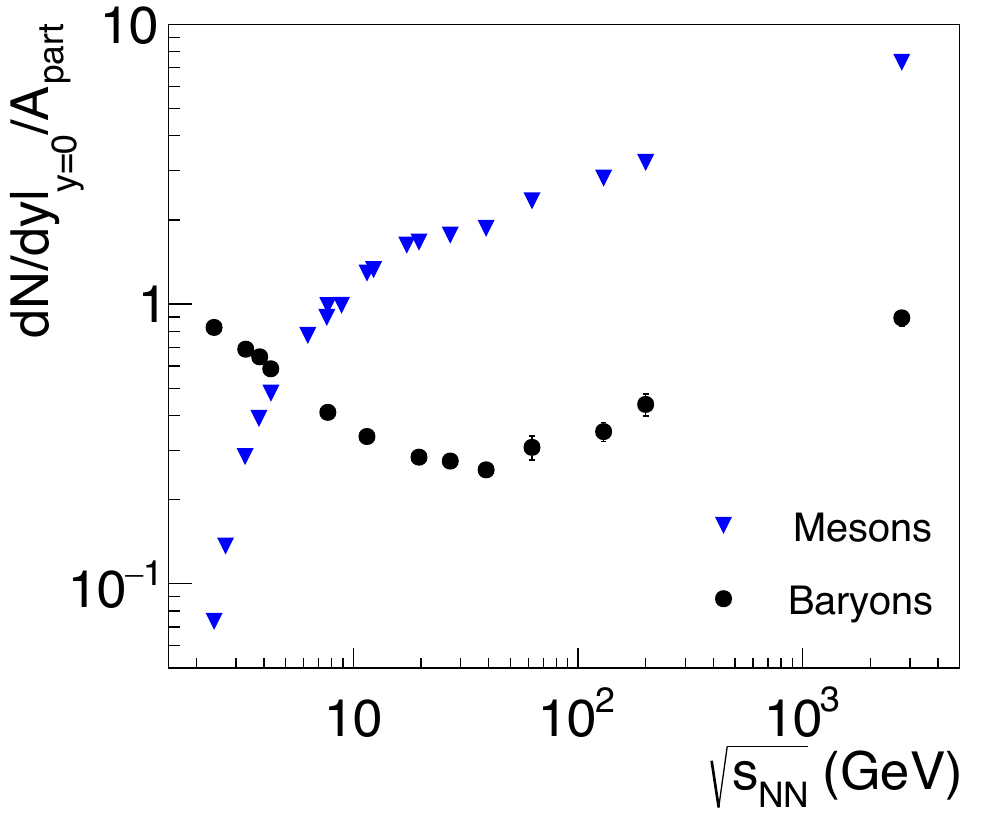}
\caption[Energy excitation functions of meson and baryon yields]{Energy excitations function of meson (blue triangles) and baryon (black dots) yields emitted at mid-rapidity in central Au+Au (Pb+Pb) collisions.}
\label{mesonbaryonsqrts}
\end{figure} 

\subsection{The transition from baryon- to meson dominated matter} 
\label{sect:baryon-meson}
Based on the measured yields of identified hadrons, we construct the excitation functions of baryons and mesons and thereby determine the collision energy at which the system created in the interaction zone transitions from a baryon- to a meson-dominated regime.
The meson yield is dominated by the two lightest mesons, pions and kaons. 
Heavier, strongly decaying mesons predominantly cascade into these light pseudoscalar ground states. 
Consequently, the excitation function of the total meson yield can be reasonably well approximated by the sum of the $\pi$ and $K$ yields 
\cite{HADES:2017jgz,HADES:2020ver,E866:1999ktz,E866:2000dog,E-0895:2003oas,NA49:2002pzu,NA49:2007stj,STAR:2017sal,STAR:2008med,ALICE:2013mez}.
Since measurements of neutral pions and kaons are sparse compared to the charged states, their yield is calculated as $(M^-+M^+)/2=M^0$ in case of pions and $(M^-+M^+)=M^0$ in case of kaons in the following.

The corresponding excitation function of the meson yield at mid-rapidity, normalized by the number of participant nucleons $A_{\text{part}}$, is shown in Fig.~\ref{mesonbaryonsqrts} (blue triangles). 
The yield rises continuously and remarkably smoothly, from below $0.1$ per participant at a few~GeV to nearly $10$ at $\sqrt{s_{NN}}=5$~TeV. 
While the increase is steep at low $\sqrt{s_{NN}}$, the slope flattens between about $5$ and $10$~GeV.

In contrast to mesons, the baryonic composition changes more strongly with $\sqrt{s_{NN}}$. 
At all energies we infer the neutron yield by multiplying the measured proton yield by the initial neutron-to-proton ratio. 
For $\sqrt{s_{NN}}\lesssim 5$~GeV we also account for baryons bound in light nuclei \footnote{Which reduces the amount of data points, due to limited availability of mid-rapidity yields of light nuclei.} (up to ${}^3\mathrm{He}$) at $\sqrt{s_{NN}} = 2.4$~GeV. 
For $\sqrt{s_{NN}}\gtrsim 3$~GeV we include single-strange hyperons, mainly $\Lambda$ and $\Sigma^{0,\pm}$; when only mid-rapidity data on $(\Lambda+\Sigma^0)$ are available, we estimate the inclusive single-strange hyperon yield as $M_Y \simeq 2\,M_{\Lambda+\Sigma^0}$, assuming isospin symmetry within the $\Sigma$ triplet \cite{E895:2001yfr,NA49:2008ysv,NA49:2010lhg,STAR:2017sal,ALICE:2013cdo}. 
For $\sqrt{s_{NN}}\gtrsim 10$~GeV, antibaryon yields (antiprotons and antihyperons) become non-negligible and are included analogously.

The mid-rapidity baryon yield (black circles in the left panel of Fig.~\ref{mesonbaryonsqrts}) varies smoothly with energy: it decreases from $\approx 0.9$ at $\sqrt{s_{NN}}=2.4$~GeV to a minimum of $\approx 0.3$ at $\sqrt{s_{NN}}\approx 40$~GeV, and then increases again to $\approx 0.8$ at $\sqrt{s_{NN}}=5$~TeV. 
The meson yield overtakes the baryon yield for $\sqrt{s_{NN}}\gtrsim 5$~GeV. 
This reflects the decreasing stopping of initial baryons with rising energy together with the rapidly growing production of new hadrons, which is dominated by mesons. 
Consequently, the system formed around mid-rapidity evolves from being (net-)baryon-dominated to meson-dominated with increasing $\sqrt{s_{NN}}$.

%
%
\subsection{Quantitative relation of stopping and particle production}

\begin{figure}[tb]
  \centering
  \includegraphics[width=0.49\textwidth]{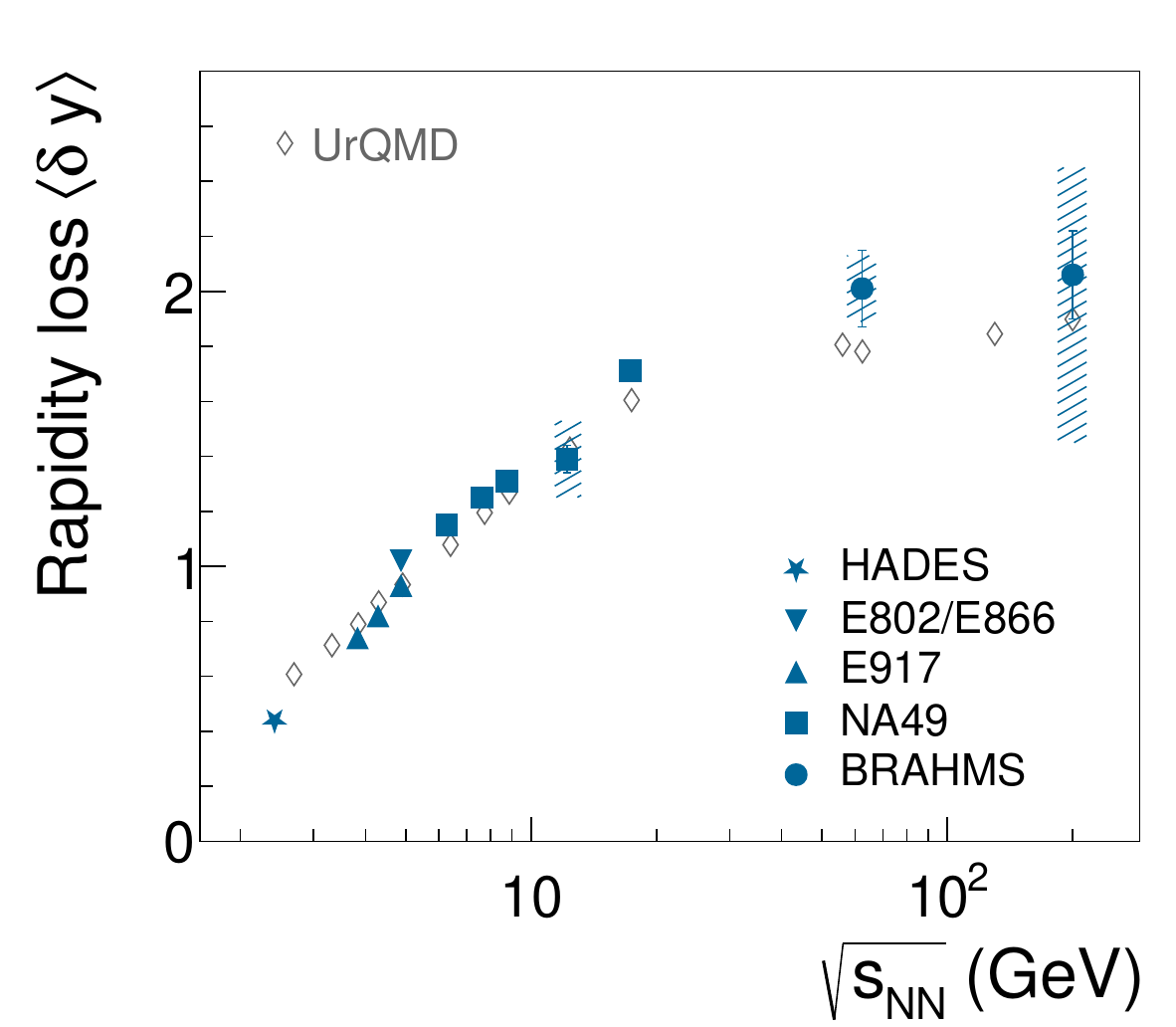}
  \includegraphics[width=0.49\textwidth]{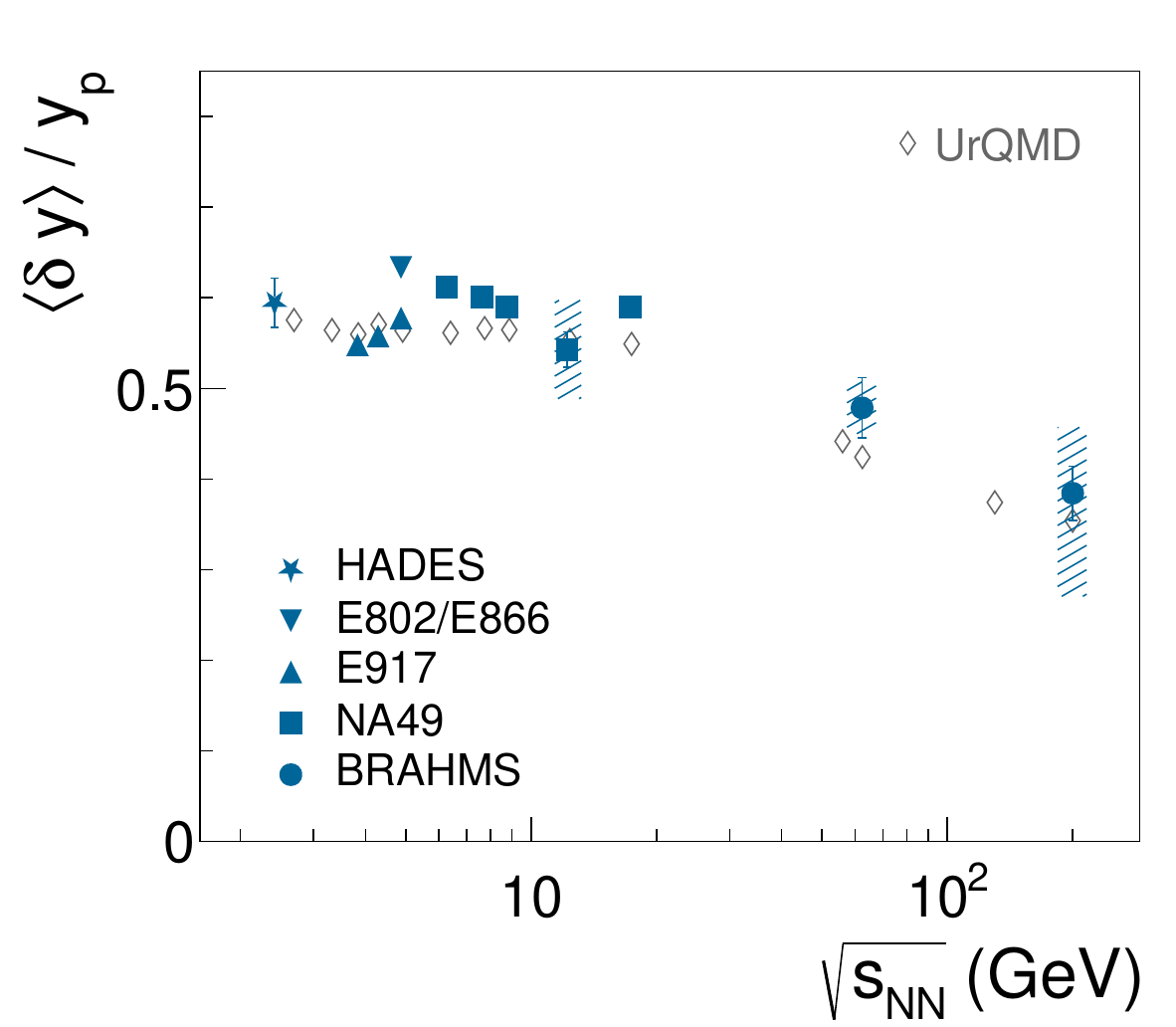}
  \caption[Stopping.]{
    Left: the average rapidity loss $\langle\delta y\rangle$ of the net-baryons as a function of the center-of-mass energy $\sqrt{s_{NN}}$, as extracted from available measurements of net-baryon distributions \cite{E802:1999hit,E917:2000spt,Videbaek:1995mf,Blume:2007kw,NA49:1998gaz,NA49:2008ysv,NA49:2010lhg,BRAHMS:2009wlg,BRAHMS:2003wwg}.  The hatched boxes depict the systematic uncertainties due to the extrapolations.  Also shown as open gray symbols are predictions by the UrQMD model \cite{Petersen:2006mp}.
    Right: the average rapidity loss of the net-baryons divided by the projectile rapidity $y_p$. 
  }
  \label{fig:stopping}
\end{figure}

Using the measured net-baryon rapidity distributions $\textrm{d}N_{(B-\bar{B})}/\textrm{d}y$, as e.g. shown in Fig.~\ref{fig:net-baryons}, the average rapidity loss $\langle\delta_y\rangle$ can be derived:
\begin{equation}
  \langle \delta y \rangle = y_{p} 
    - \frac{2}{N_{part}} 
      \int_{0}^{y_{p}} y \: 
      \frac{\textrm{d}N_{(B-\bar{B})}}
           {\textrm{d}y} \: 
      \textrm{d}y,
\end{equation}
where $y_p$ is the projectile rapidity and $N_{part}$ the number of participating nucleons.  As the measurements of $\textrm{d}N_{(B-\bar{B})}/\textrm{d}y$ usually do not cover the complete longitudinal phase space, extrapolations based on different assumptions have to be made.  For Au+Au collisions at $\sqrt{s_{NN}} = 200$~GeV \cite{BRAHMS:2003wwg}, for instance, a six order symmetric polynomial and a symmetric sum of two Gaussians in momentum space, converted to rapidity space, are used to derive an upper and lower limit on $\langle\delta_y\rangle$.  At lower energies (HADES, E917, NA49) the symmetric sum of two Gaussians in rapidity space or symmetric polynomials generally provide good descriptions \cite{E917:2000spt,Blume:2007kw} and their differences mainly define the uncertainties of the extracted $\langle\delta_y\rangle$ values.  The evolution of the thus resulting $\langle\delta y\rangle$ with center-of-mass energy is summarized in the left panel of Fig.~\ref{fig:stopping}.  A continuous rise of the rapidity loss is observable, which tends to saturate for high $\sqrt{s_{NN}}$.  The right panel presents the same quantity, only divided by the projectile rapidity $y_p$.  While at lower energies, between SIS18 and top-SPS, approximately constant values of $\approx 0.6$ are observed for $\langle\delta y\rangle / y_p$, the relative rapidity loss drops to $\approx 0.4$ at $\sqrt{s_{NN}} = 200$~GeV.  Consequently, the stopping power at higher center-of-mass energies is not any more sufficient to transport the initial baryon number completely towards mid-rapidity, as the rapidity gap between projectile and target, $\Delta y = y_p - y_t = \tanh^{-1} (p_{z}^{beam}/E^{beam})$, increases faster than the average rapidity loss.  The data points shown in Fig.~\ref{fig:stopping}  agree with a prediction by the UrQMD model \cite{Petersen:2006mp}, demonstrating that global properties like $\langle\delta_y\rangle$ can be well described within a transport model approach.

\begin{figure}[tb]
  \centering
  \includegraphics[width=0.49\textwidth]{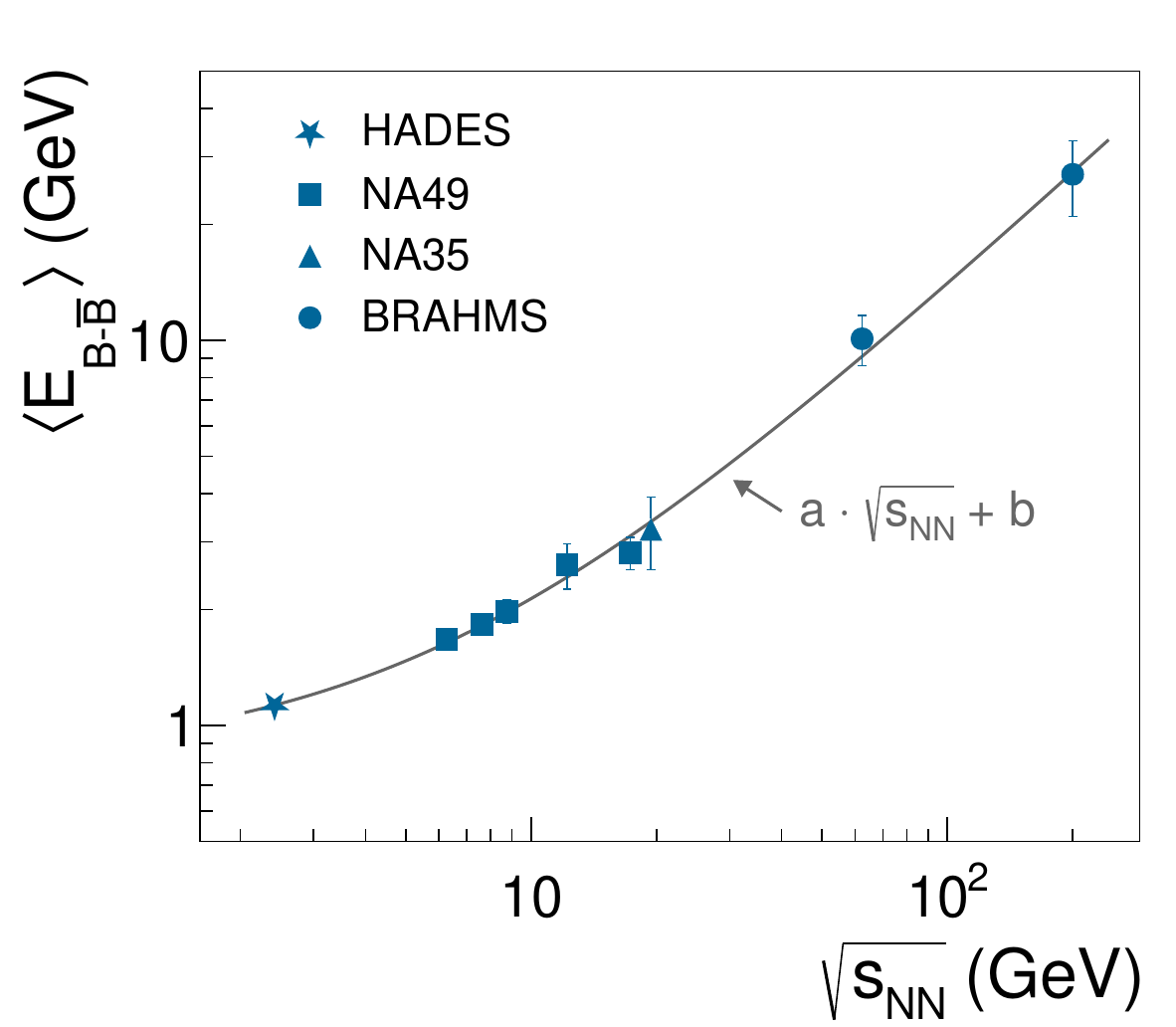}
  \includegraphics[width=0.49\textwidth]{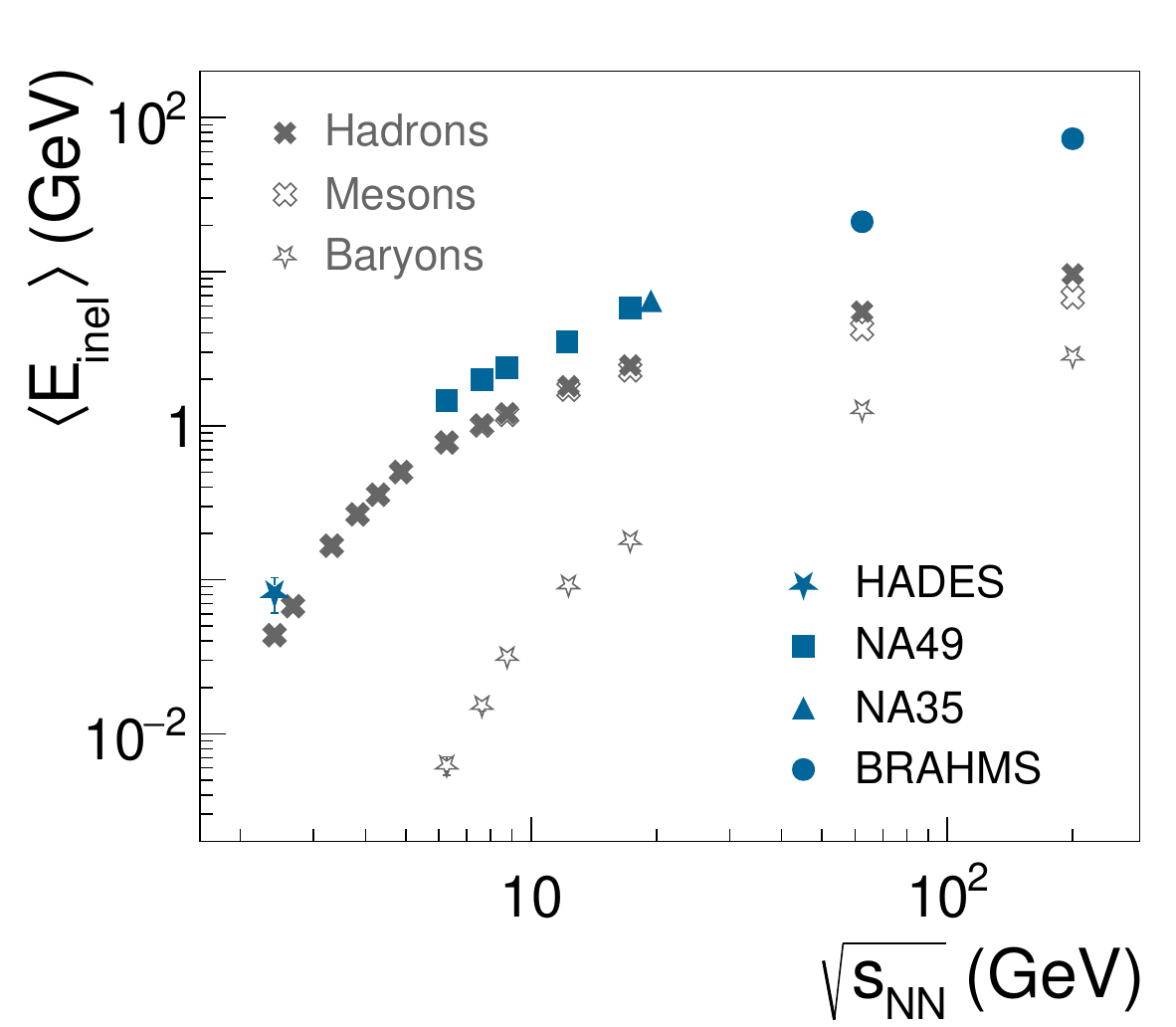}
  \caption[Energies.]{
    Left: the average energy per net-baryon $\langle E_{B- \bar{B}} \rangle$ as a function of the center-of-mass energy $\sqrt{s_{NN}}$ \cite{Blume:2007kw,NA49:1998gaz,NA49:2008ysv,NA49:2010lhg,BRAHMS:2009wlg,BRAHMS:2003wwg} for central Au+Au(Pb+Pb) collisions.  In addition, also the value extracted by the NA35 collaboration for central S+S collisions at 200~$A$GeV is shown \cite{NA35:1994adm}.  The data points are fitted with a function $f = a \cdot \sqrt{s_{NN}} + b$, yielding the parameters $a = 0.13 \pm 0.01$ and $b = 0.81 \pm 0.03$~GeV.
    Right: the average inelastic energy per net-baryon $\langle E_{inel} \rangle$.  Also shown as gray symbols is an estimate on the amount of inelastic energy used for the creation of hadronic mass, calculated for mesons (open crosses), baryons (open stars) and their sum (hadrons, filled crosses), see text for details.
  }
  \label{fig:enetbaryon}
\end{figure}

Using $\textrm{d}N_{(B-\bar{B})}/\textrm{d}y$ and the measured mean transverse mass\footnote{$m_t = \sqrt{p_t^2 + m_0^2}$, where $m_0$ is the rest mass of a given particle.} $\langle m_{t} \rangle$ of protons the average energy per net-baryon can be estimated in a next step:
\begin{equation}
  \langle E_{B-\bar{B}} \rangle 
     = \frac{1}{N_{(B-\bar{B})}}
       \int_{-y_{p}}^{y_{p}} \langle m_{t} \rangle \:
       \frac{\textrm{d}N_{(B-\bar{B})}}{\textrm{d}y} 
       \: \cosh y \: \textrm{d}y \;.
\end{equation}
As $\langle m_t \rangle$ often is only measured at mid-rapidity, assumptions on its dependence on rapidity have to be made in these cases.  E.g. in \cite{Blume:2007kw,BRAHMS:2009wlg,BRAHMS:2003wwg} a linear and a Gaussian shaped evolution between the $\langle m_t \rangle$ value at mid-rapidity and the proton mass $m_p$ at projectile and target rapidities are used and the differences between the extrapolations are included in the uncertainties.  The same procedure is here also used for the preliminary HADES data.  The left panel of Fig.~\ref{fig:enetbaryon} summarizes the thus extracted $\langle E_{B-\bar{B}} \rangle$ values for different experiments\footnote{For the available AGS data such an analysis was not performed in \cite{E802:1999hit,E917:2000spt,Videbaek:1995mf} and therefore no values are shown here.}.  Their rise with $\sqrt{s_{NN}}$ can be very well described by a linear function, as demonstrated by the fit shown as well in Fig.~\ref{fig:enetbaryon}.

The average inelastic energy, defined for the nucleon-nucleon system, i.e. the kinetic energy of an incoming nucleon that is actually converted into particle production and dynamics, as well as radiation, is given by the difference between the initial center-of-mass energy per nucleon and the average energy per net-baryon:
\begin{equation}
  \langle E_{inel} \rangle = \frac{\sqrt{s_{NN}}}{2} - \langle E_{B-\bar{B}} \rangle
\end{equation}
The right panel of Fig.~\ref{fig:enetbaryon} displays the dependence of $\langle E_{inel} \rangle$ on center-of-mass energy, which exhibits a continuous rise proportional to $\sqrt{s_{NN}}$.  In order to quantify the amount of inelastic energy that is used for the generation of hadronic mass only (i.e. for static hadrons), the measured total hadron yields are multiplied by their respective rest masses.  The total meson yields are constructed in the same way as outlined above (see Sect.~\ref{sect:baryon-meson}), however, here the 4$\pi$ integrated yields are used instead of mid-rapidity $\textrm{d}N/\textrm{d}y$ in order to capture the total particle production (data are taken from \cite{E-0895:2003oas,E-802:1998xum,E866:1999ktz,E866:2000dog,NA49:2007stj,NA49:2002pzu,Blume:2008zza,NA49:2010lhg,NA49:2008ysv,BRAHMS:2009acd,BRAHMS:2004dwr,BRAHMS:2009wlg,BRAHMS:2003wwg,STAR:2008med,STAR:2006egk,STAR:2010yyv}).  The mesonic mass generation constitutes a large fraction of the available inelastic energy and rises, at least in the energy region below SPS, in a similar way with $\sqrt{s_{NN}}$ as $\langle E_{inel} \rangle$.  The baryonic contribution to the newly generated mass, i.e. not the mass of the initially present baryons, can be estimated by multiplying the measured total antiproton yields by a factor four to account for produced protons and (anti-)neutrons.  For higher energies also the measured total $\bar{\Lambda} + \bar{\Sigma}^0$ yields, multiplied by a factor four to account for the produced $\Lambda^+ + \Sigma^0$ and the unmeasured $\Sigma^\pm$ and $\bar{\Sigma}^\mp$, are added.  The baryons contribute to a much lesser extend than the mesons in the studied energy range and only above the p$\bar{\textrm{p}}$-threshold.

\begin{figure}[tb]
  \centering
  \includegraphics[width=0.49\textwidth]{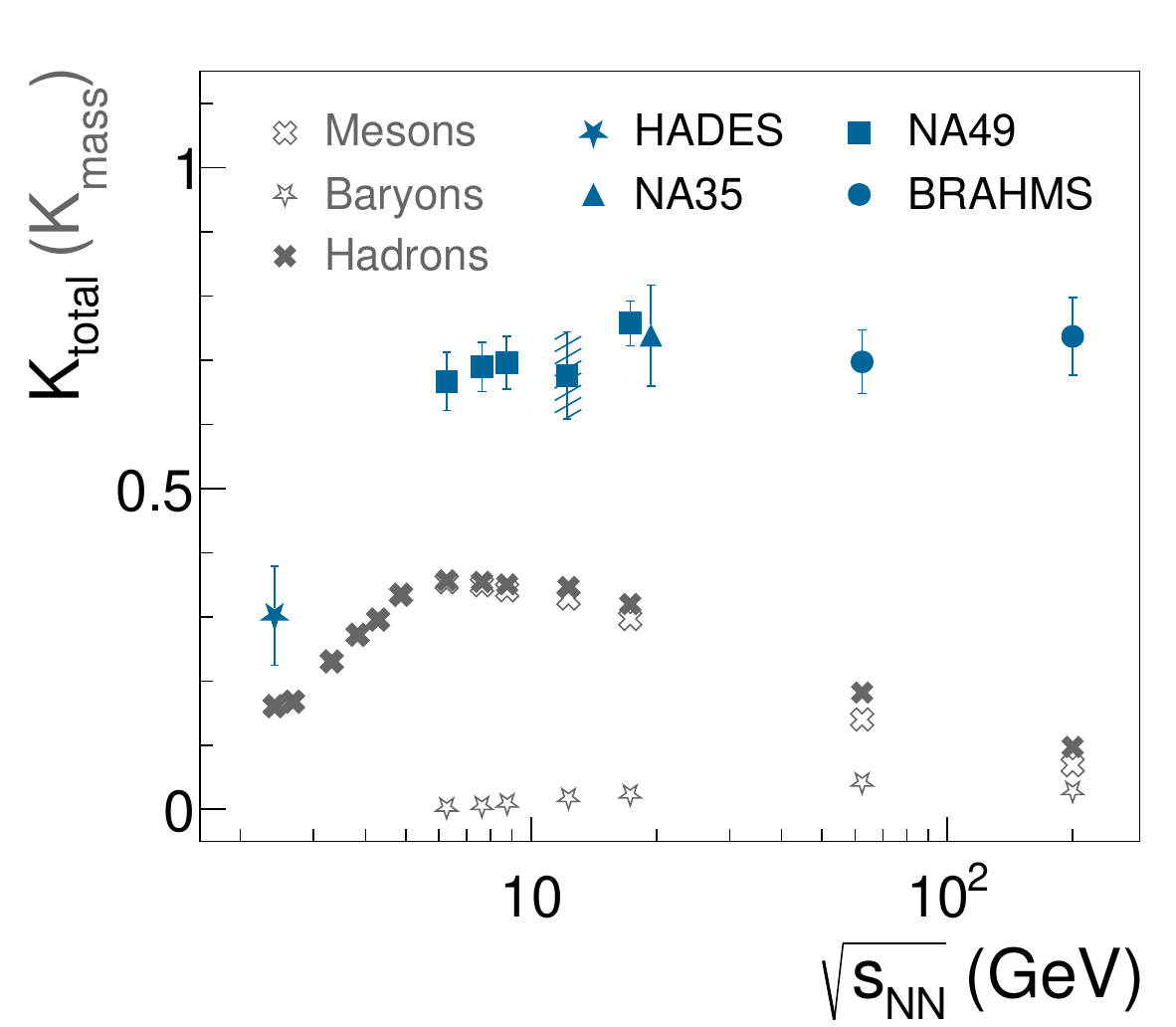}
  \includegraphics[width=0.49\textwidth]{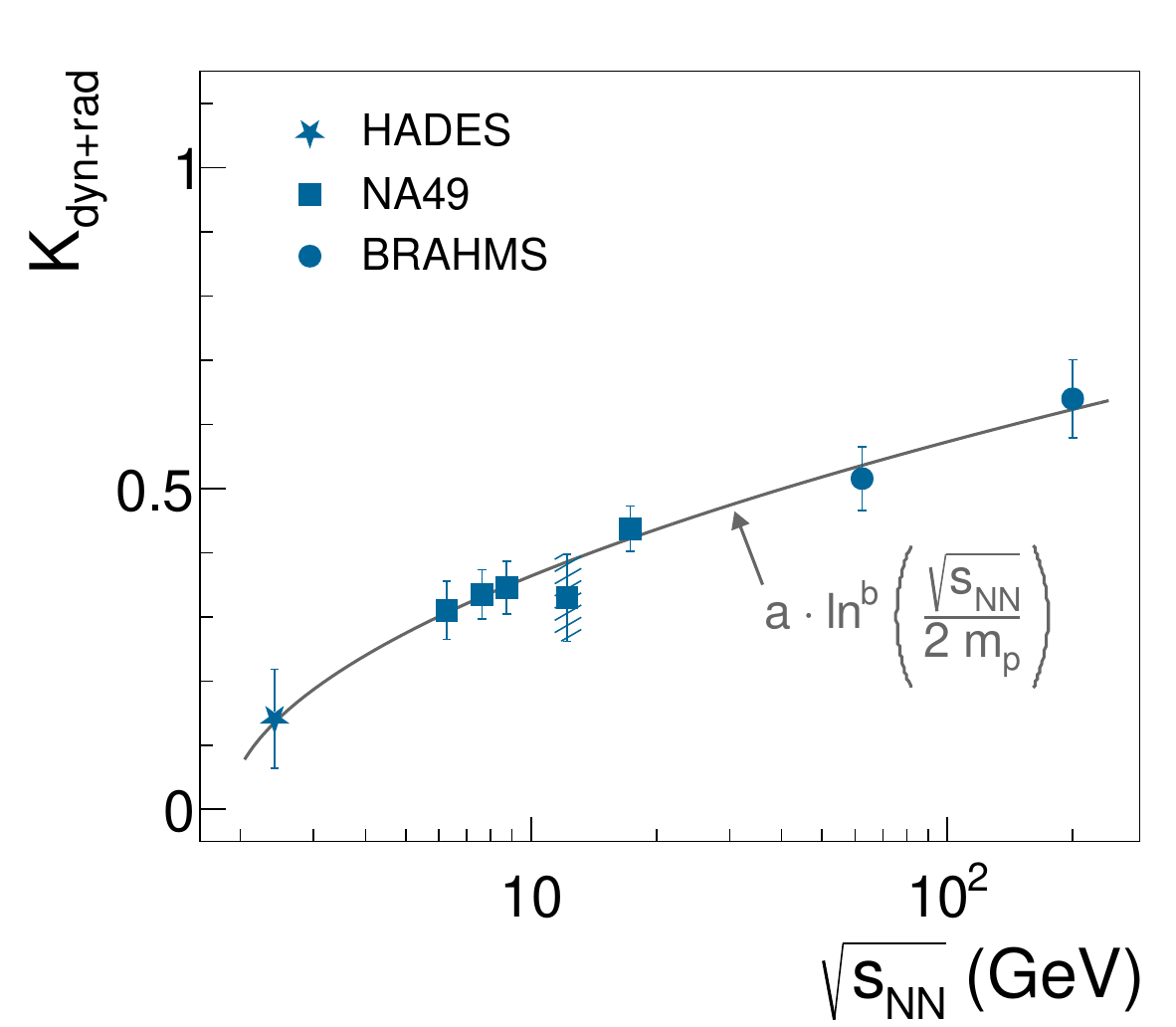}
  \caption[Inelasticity.]{
    Left: the total inelasticity $K_{total}$ as a function of the center-of-mass energy $\sqrt{s_{NN}}$ (colored, filled symbols). The fraction $K_{mass}$ that is used for hadronic mass generation is also shown as gray symbols for mesons (open crosses), baryons (open stars) and their sum (hadrons, filled crosses).
    Right: the fraction of the inelasticity going into the dynamics of newly produced particles and into radiation $K_{dyn+rad}$.  The data points are fitted with a function $f = a \cdot \ln^b(\sqrt{s_{NN}} / (2 \, m_p))$, yielding the parameters $a = 0.28 \pm 0.02$ and $b = 0.52 \pm 0.09$.
  }
  \label{fig:inelasticity}
\end{figure}

\begin{figure}[tb]
  \centering
  \includegraphics[width=0.49\textwidth]{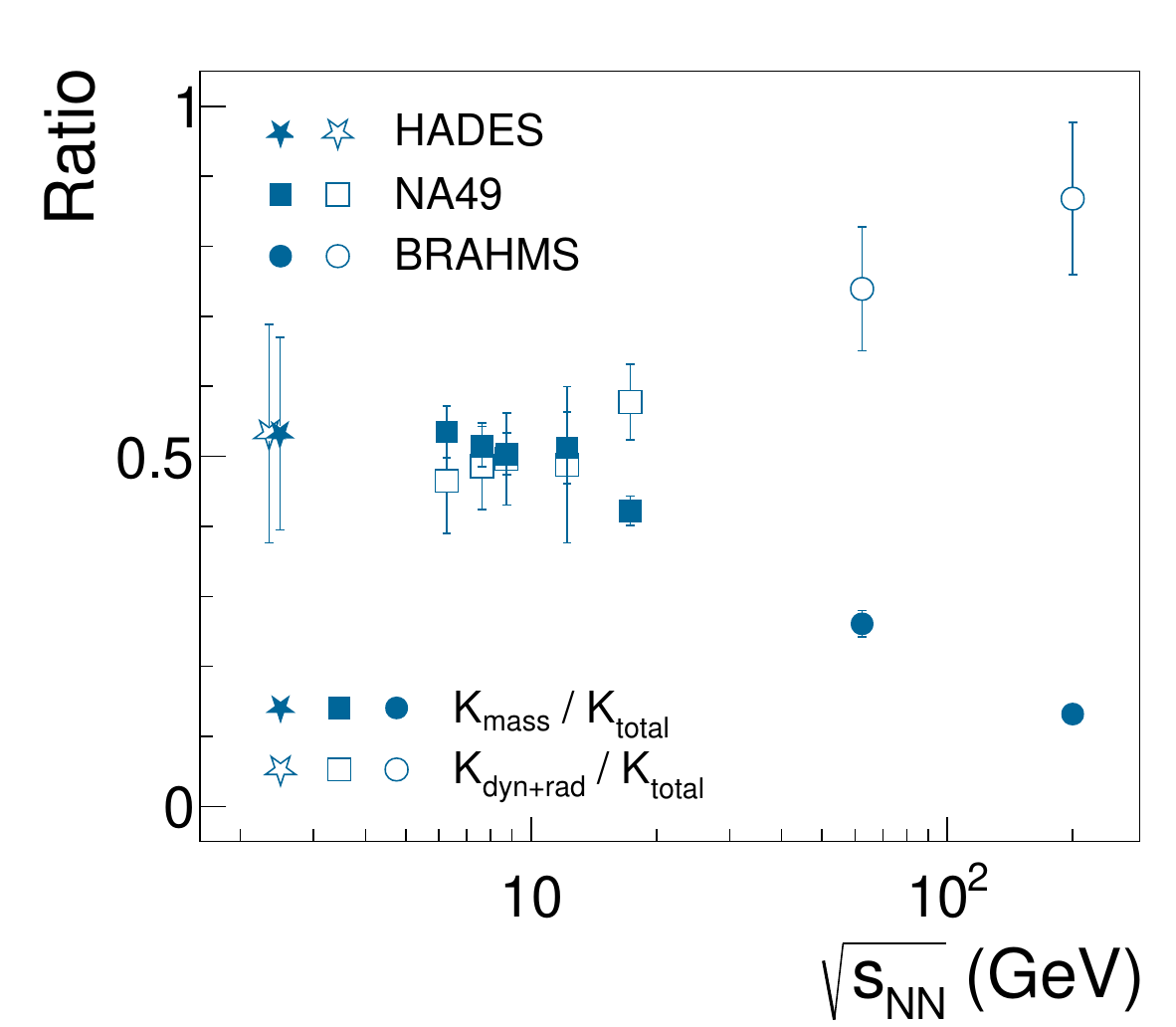}
  \includegraphics[width=0.49\textwidth]{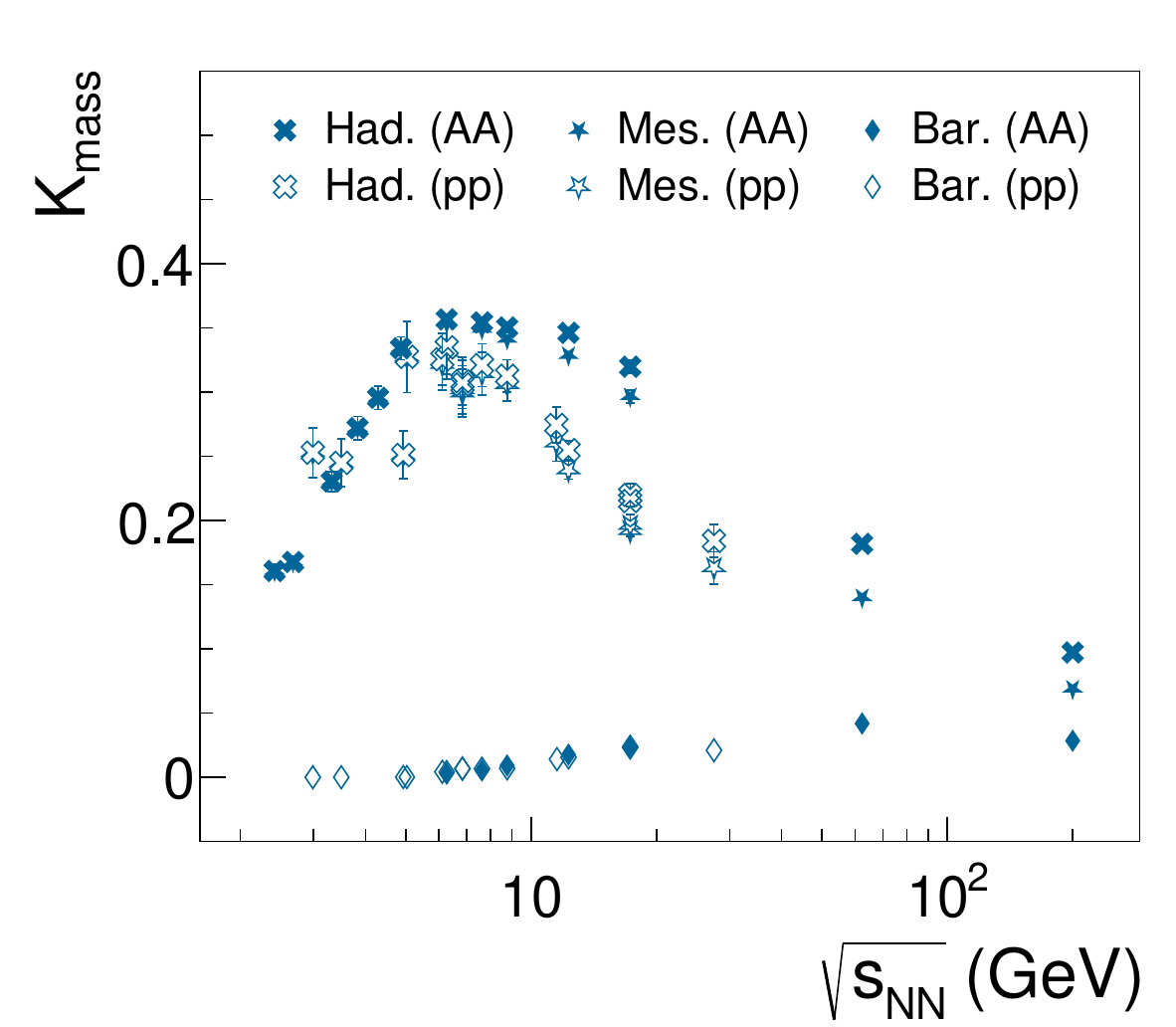}
  \caption[Energies.]{
    Left: the ratios $K_{mass} / K_{total}$ (filled symbols) and $K_{dyn+rad} / K_{total}$ as a function of the center-of-mass energy $\sqrt{s_{NN}}$.
    Right: comparison of the fraction of the inelastic energy that is used for hadronic mass generation $K_{mass}$ for AA and pp collisions.  Shown are the values for the sum of produced mesons and baryons, as well as the sum.
  }
  \label{fig:ratios+generated_mass}
\end{figure}

Finally, the total inelasticity $K_{total}$ is calculated using the above defined $\langle E_{inel} \rangle$ as:
\begin{equation}
  K_{total} = 2 \: \langle E_{inel} \rangle \,/ \,(\sqrt{s_{NN}} - 2 \, m_{p}) \, .
\end{equation}
It quantifies the fraction of the initial kinetic energy available per nucleon-nucleon collision in the center-of-mass system, that is converted into inelastic reactions.  As shown in the left panel of Fig.~\ref{fig:inelasticity}, $K_{total}$ rises from a value of around 0.3 at $\sqrt{s_{NN}} = 2.4$~GeV to an approximately energy independent value of $K_{total} \approx 0.7 - 0.8$ above $\sqrt{s_{NN}} \approx 10$~GeV.  The fraction that is used for mass generation $K_{mass}$ is defined in a similar way 
\begin{equation}
  K_{mass} = \left. \left( \sum_i \, \frac{2}{N_{part}} \, N_i \, m_i \right) 
           \right/ \,(\sqrt{s_{NN}} - 2 \, m_{p}) \, ,
\end{equation}
where $N_i$ are the total yields and $m_i$ the rest masses of the particles discussed above, and is also shown in the left panel of Fig.~\ref{fig:inelasticity} (gray symbols).  $K_{mass}$ rises until $\sqrt{s_{NN}} \approx 6$~GeV to a maximum value of $\sim 0.35$ and continuously drops from thereon.  This means that from this center-of-mass energy on-wards particle generation becomes less and less favored than longitudinal and transverse dynamics, as well as the emission of radiation.  This is also visible in the right panel of Fig.~\ref{fig:inelasticity}, which shows that the difference $K_{dyn+rad} = K_{total} - K_{mass}$ is continuously rising by more than a factor three from $K_{dyn+rad} \approx 0.2$ at SIS18 energies to $K_{dyn+rad} \approx 0.65$ at top RHIC energy.  This dependence can be very well described by a function proportional to $\ln^{1/2}(\sqrt{s_{NN}})$, as shown by the solid line in Fig.~\ref{fig:inelasticity}, right.  From the above observations it follows that until $\sqrt{s_{NN}} \approx 10$~GeV the inelasticity is shared in equal amount be particle production ($K_{mass}$) and dynamics ($K_{dyn+rad}$), while above this energy dynamics and radiation are more favored, reaching up to $~90$~\% of the total inelasticity at $\sqrt{s_{NN}} = 200$~GeV, while $K_{mass}$ drops to almost $10$~\% of $K_{total}$ (see left panel of Fig.~\ref{fig:ratios+generated_mass}).

From the above compilations a consistent, overall picture of a continuous evolution of and stopping and inelasticity emerges.  While at low energies (SIS18) the available kinetic energy of the initial nucleons is mainly transformed into the production of mesons with relatively small kinetic energies, at higher energies these produced particles acquire more and more transverse and longitudinal energy.  The rise of this dynamic fraction with center-of-mass energy compensates the decrease of the fraction due to mass generation in a way that the inelasticity remains roughly constant above $\sqrt{s_{NN}} \approx 6$~GeV.  An extension of this study to even higher energies (LHC) is right now not possible, since the available experiments are not able to cover the full longitudinal phase space.  However, at low energies (e.g. FAIR) these kind of measurements will be feasible and should, on one side, be able to provide data of much higher precision as currently available and, on the other side, will fill the gaps existing in the energy region below SPS.  In addition, it would be highly desirable to collect corresponding reference data on pp-collisions, in order to be able to perform a similar analysis for elementary reactions.  As shown in the right panel of Fig.~\ref{fig:ratios+generated_mass} there is already a quite comprehensive set of data on particle production in pp-collisions \cite{QCDatFAIR}, which, however, lacks accuracy and completeness in the low energy region.  As the comparison shown demonstrates, there is a clear difference in the energy dependence of meson production between pp and AA collisions, as has already been pointed out some time ago \cite{Gazdzicki:1995ze,NA49:2002pzu}.  How the $\sqrt{s}$ evolution of stopping and $K_{total}$ in pp-collisions compares to AA is up-to-now unexplored, as data on net-proton distributions in pp-collisions are scarce and currently do not allow for a systematic investigation.  

\section{Conclusions}
We have compiled excitation functions of the approximated baryon and meson rapidity
densities at mid-rapidity, $\mathrm{d}N/\mathrm{d}y|_{y\approx 0}$, for relativistic heavy-ion
collisions from SIS18 up to the LHC. At low beam energies (SIS18/AGS), strong nuclear
stopping leads to large net-baryon densities at mid-rapidity with sizable nuclear-cluster production; the net-baryon $\mathrm{d}N/\mathrm{d}y$ shows a strongly peaked distribution around mid-rapidity and clearly dominates over the meson yields. With increasing
$\sqrt{s_{NN}}$, the net-baryon $\mathrm{d}N/\mathrm{d}y|_{y\approx 0}$ decreases
monotonically and its shape gradually evolves from a single central peak towards a
double-humped, more transparent configuration, while the pion and kaon
$\mathrm{d}N/\mathrm{d}y$ at mid-rapidity rise rapidly. In the AGS–low SPS regime this
leads to a qualitative switch from a baryon-dominated to a meson-dominated
fireball, as reflected by meson-to-baryon yield ratios at mid-rapidity crossing
unity at $\sqrt{s_{NN}}\approx 5$~GeV. At LHC energies, the mid-rapidity region is net-baryon
free, and hadron production is governed by copious meson emission. 
We find a smooth $\sqrt{s_{NN}}$ evolution of the stopping power, as quantified by the first moment of the net-baryon distributions $\langle \delta y \rangle$.  The relative stopping power $\langle \delta y \rangle / y_p$ rises continuously with increasing center-of-mass energy.  The inelasticity, on the other side, exhibits a rapid increase between SIS18 and SPS energies and appears to be constant from thereon.  While at low energies up to $\sim 10$~GeV the available energy seems to be shared by equal amount between the production of new particles and the dynamics of the system, as well as radiation, the latter part starts to dominates at higher energies.  As the current data situation is still incomplete and/or dominated by large systematic uncertainties, new measurements over large regions of phase space are needed in order to come to more firm conclusions.  The CBM experiment at FAIR will be able to provide relevant data with high statistics and low systematic uncertainties and thus will allow for a precise study of evolution of stopping and particle production in the low energy region.  It will also be important to collect pp reference data at the same energies, in order to be able to discern any differences between elementary and heavy-ion reactions and thus to identify possible signs of the deconfinement phase transition.

\vspace{6pt}

\authorcontributions{}

\funding{}

\dataavailability{}

\acknowledgments{}

\reftitle{References}

\bibliography{bibnew2_no_doi}

\end{document}